\documentclass[reprint,nofootinbib,superscriptaddress,amsmath,amssymb,aps]{revtex4-1}
\usepackage{graphicx}
\usepackage{rotating}
\usepackage{dcolumn}
\usepackage{bm}
\usepackage{color}
\usepackage{mathptmx, textcomp}
\usepackage[latin1]{inputenc}
\usepackage{braket}
\usepackage[colorlinks,linkcolor=magenta,anchorcolor=cyan,citecolor=blue]{hyperref}
\usepackage[multiple]{footmisc}
\newcommand{\fig}[1]{Fig.\ref{#1}}
\def\be{\begin{equation}}
\def\ee{\end{equation}}
\def\ba{\begin{eqnarray}}
\def\ea{\end{eqnarray}}

\def\nn{\nonumber}
\def\lf{\left}
\def\rt{\right}

\newcommand{\eq}[1]{(\ref{#1})}

\def\n{\nonumber}\def\lf{\left}\def\rt{\right}\def\q{\theta}   \def\y {\psi}   \def\p {\pi}  \def\s {\sigma} \def\d {\delta} \def\f {\phi} \def\g {\gamma} \def\h {\eta}  \def\k {\kappa} \def\l {\lambda}  \def\x {\xi} \def\c {\chi}   \def\m {\mu} \def\pd {\partial}\def\p {\pi} \def \inf {\infty}  
\def\Q{\Theta} \def\W{\Omega}     \def\S {\Sigma} \def\D {\Delta} \def\F {\Phi}  \def\L {\Lambda}    \def\grad{\nabla}\def\.{\cdot}
\def\math {\mathcal}
\begin{document}

\title{Holographic complexity of charged Taub-NUT-AdS black holes}
\author{Jie Jiang}
\email{jiejiang@mail.bnu.edu.cn}
\affiliation{Department of Physics, Beijing Normal University, Beijing, 100875, China}
\author{Banglin Deng}
\email{bldeng@aliyun.com}
\affiliation{Department of Applied Physics, College of Geophysics, Chengdu University of Technology, Chengdu 610059, Sichuan, China}
\author{Xiao-Wei Li}
\email{lixiaowei16@mails.ucas.ac.cn}
\affiliation{School of Physics, University of Chinese Academy of Sciences, Beijing 100049, China}
\date{\today}

\begin{abstract}
In this paper, we investigate the holographic complexity in the charged Taub-NUT-AdS black holes with Misner strings present in the Einstein-Maxwell gravity. We show that differing from the normal black holes, where the late-time complexity growth rate is only determined by the quantities at outer and inner ``Reissner-Nordstrom''-type (RN-type) horizons, here the quantities (the Misner potential and Misner charge) related to the Misner strings also play an important role in CA complexity.  Similar to the case of the normal electromagnetic black hole, the late-time rate for the original CA conjecture is independent on the magnetic charges. However, disparate with common results of the dyonic solutions, the electric charge appeared here is the total charge of this black hole. Besides, we found that the result in this original CA conjecture also violates the electromagnetic duality. And this duality can be restored by adding the Maxwell boundary term with the proportional constant $\g=1/2$. In this case, the late-time rate is sensitive to the magnetic charge. Moreover, we also found that the additional term only changes the proportion between the electric and magnetic charges, and it does not affect the Misner term appeared in the late-time rate. Finally, we studied the time-dependence of the complexity growth rate and found that they share similar behaviors with that in RN-AdS black holes.
\end{abstract}
\maketitle
\section{Introduction}
In recent years, there has been a growing interest in the topic of ``quantum complexity", which is defined as the minimum number of gates required to obtain a target state starting from a reference state \cite{L.Susskind,2}.  From the holographic viewpoint, Brown $et\,al.$ suggested that the quantum complexity of the state in the boundary theory is dual to some bulk gravitational quantities
which are called ``holographic complexity". Then, the two conjectures, ``complexity equals volume" (CV) \cite{L.Susskind,D.Stanford} and ``complexity equals action" (CA) \cite{BrL,BrD}, were proposed. They aroused widespread attention of researchers to both holographic complexity and circuit complexity in quantum field theory, $e.g.$ \cite{Jiang:2019pgc,Bhattacharyya:2018bbv,Ali:2018fcz,Ali:2018aon,HosseiniMansoori:2018gdu,Liu:2019smx,Jiang:2019qea,A4,Pan,A2,Guo:2017rul,WYY,Jiang1,
A51,A11,A12,A13,A14,A15,A16,A17,A18,A19,A20,A23,A24,A25,A26,A27,A28,A29,A30,A31,A32,A33,A34,A35,A36,
Fan:2019mbp,Goto:2018iay,JiangL,Nally:2019rnw,Jiang3,Chapman1,Chapman2,Jiang4,SZ,Roberts:2014isa,A37,A38,A39,Jiang1,A2,A3,A4,A5,A6,A7,A8,A9,A10}.

In present work, we only focus on the CA conjecture, which states that the quantum complexity of a particular state $|\y(t_L,t_R)\rangle$ on the boundary is given by
\ba\label{CA}
C_A\lf(|\y(t_L,t_R)\rangle\rt)\equiv\frac{I_\text{WDW}}{\p\hbar}\,.
\ea
Here $I_\text{WDW}$ is the on-shell action in the corresponding Wheeler-DeWitt (WDW) patch, which is enclosed by the past and future light sheets sent into the bulk spacetime from the timeslices $t_L$ and $t_R$. As argued in \cite{BrL}, there is a bound of the complexity growth rate  at the late time
\ba
\dot{C}\leq \frac{2 M}{\p \hbar}\,,
\ea
which may be thought as the Lloyd's bound of the black hole system \cite{SLloyd}.

In the previous works, the CA conjecture in the normal black holes for variety of gravitational theories have been investigated. In Ref. \cite{Jiang1}, based on the Noether charge formalism introduced by Iyer and Wald, a general formalism has been obtained to describe the late-time CA complexity growth rate in a multiple-horizon black hole for a general F(Riemann) gravity, i.e.,
\ba\label{dIdt}\begin{aligned}
\lim_{t\to\inf}\frac{dC_A}{dt}=\frac{1}{\p\hbar}\left[\W^{(\m)}J_{(\m)}+\L_{\inf}[\x]-\L_{\math{C}}[\x]\right]^{-}_{+}\,,
\end{aligned}\ea
where $\W^{(\m)}$ and $J_{(\m)}$ are the angular velocity and angular momentum associated with the Killing horizon respectively, $\L_\math{C}$ denotes some charges related to the matter field, and the index $\{\pm\}$ presents the quantities evaluated at the ``outer'' or ``inner'' horizon. This result implies that the late-time rate only depends on the quantities on the horizons. And this can be understood since the boundary of the WDW patch is only bounded by the spheres $\S_\pm$ on the outer/inner horizon and $S_\inf$ at asymptotic infinity. However, there are also some black holes with different topology than normal black hole, such as the Taub-NuT-AdS black hole with the Misner strings \cite{Taub, Newman}. How is this nontrivial topology reflected in the holographic complexity? For this reason, the purpose of this paper is to study the effect of the non-trivial topology of the spacetime on the holographic complexity in the CA picture.

As one of the most interesting solutions of general relativity, Taub-NUT spacetime \cite{Taub, Newman} has engendered lots of investigations since birth. In this spacetime, apart from the RN-type horizons, there also exist two Misner string singularities on the north and south pole axes due to the existence of NUT parameter $n$. This Misner string is also a Killing horizon of this black hole and it will actually change the topology of the spacetime geometry, especially the late-time boundaries of the WDW patch. As illustrated by Refs. \cite{DK1,DK2}, there exist two quantities to describe the Misner string: the Misner potential $\y$ and charge $N$, which is introduced to make the thermodynamics of Taub-NUT black hole has the expected features. In the rest of this paper, we would like to evaluate the CA complexity growth rate and discuss its corrections caused by this Misner string.

Our paper is organized as follows. In the next section, we briefly review the rudiments about the charged Taub-NUT-AdS spacetime and enumerate some basic thermodynamic quantities for this black hole. In Sec. \ref{3}, we evaluate the time-dependence of the complexity growth rate in the charged Taub-NUT-AdS black hole in the original CA picture and discuss the affection of the Misner string. In Sec. \ref{4}, we investigate the CA conjecture with the additional Maxwell boundary term. Finally, conclusions are presented in Sec. \ref{5}.

\section{Geometry of charged TAUB-NUT-AdS spacetime}
In this paper, we consider the charged Taub-NUT-AdS solution of the $4$-dimensional Einstein-Maxwell gravity, where the bulk action can be written as
\ba
I_\text{bulk}=\frac{1}{16\pi}\int_{M}d^4x\sqrt{-g}\lf(R-2\L-F_{ab}F^{ab}\rt)\,,
\ea
in which $\bm{F}=d\bm{A}$ is the electromagnetic strength and $R$ is the Ricci scalar of the spacetime. The equations of motion are given by
\ba\label{eom}\begin{aligned}
R_{ab}-\frac12Rg_{ab}+\L g_{ab}&=T_{ab}\,,\\
d\bm{G}=0,
\end{aligned}\ea
with the energy-momentum tensor of the Maxwell field
\begin{align}
T_{ab}=F_{ac}{F_b}^c+G_{ac}{G_b}^c \label{Tab}\,,
\end{align}
and
\begin{align}
{\bm G}=\star{\bm F}\,.
\end{align}
The electric and magnetic charges inside a $2$-dimensional surface $S^2$ can be defined as
\ba\begin{aligned}\label{QeQm}
q_e[S^2]=\frac{1}{4\pi}\int_{S^2}\bm{G}\,,\ \ \ q_m[S^2]=\frac{1}{4\pi}\int_{S^2}\bm{F}\,.
\end{aligned}\ea
According to the equations of motion \eq{eom}, the charged Taub-NUT-AdS solution can be read off \cite{Johnson2,NA}
\ba\begin{aligned}\label{A}
ds^2=&-f(r)\left(dt+2n\cos\theta\,d\phi\right)^2+\frac{dr^2}{f(r)} \\
&+\left(r^2+n^2\right)\left(d\theta^2+\sin^2\theta\,d\phi^2\right)\,,\\
{\bm A}=&-\lf[h(r)-h_0\rt]dt-2 h(r) n\cos\theta\,d\phi\,,
\end{aligned}\ea
where
\begin{align}
f(r)=\frac{r^2-2m r-n^2+4n^2 g^2+e^2}{r^2+n^2}-\frac{3n^4-6n^2r^2-r^4}{l^2(r^2+n^2)} \n
\end{align}
is the blackening factor,
\begin{align}
h(r)=\frac{e r}{r^2+n^2}+\frac{g(r^2-n^2)}{r^2+n^2}\,,
\end{align}
and $h_0$ is some arbitrary constant which reflects the gauge freedom of the electromagnetic field. Besides, $n$ and $m$ are the NUT and mass parameters, $e$ and $g$ are the electric and magnetic parameters, and $l$ is the AdS radius with $\Lambda=-3/l^2$. In this paper, we only consider the solution which has two RN-type horizons which are located on the sphere with radius of $r=r_\pm$. And we can see that in this solution, apart from the RN-type horizons, there also exist two Misner string singularities on the north and south pole axes. And this Misner string will affect the topology of this spacetime.

According to Eq. \eq{A}, we can immediately obtain electromagnetic field tensor
\ba\begin{aligned}\label{expF}
{\bm F}&=-\frac{e(n^2-r^2)+4n^2gr}{(r^2+n^2)^2}\,dr\wedge dt+\frac{2n[er+g(r^2-n^2)]}{r^2+n^2}\\
\times&\sin\theta d\theta\wedge d\phi-2n\frac{e(n^2-r^2)+4n^2gr}{(r^2+n^2)^2}\cos\theta\,dr\wedge d\phi.
\end{aligned}\ea
Then, we can instantly obtain
\ba\begin{aligned}\label{expG}
{\bm G}&=\frac{2n[g(n^2-r^2)-er]}{(r^2+n^2)^2}\,dr\wedge dt+\frac{e(r^2-n^2)-4n^2gr}{r^2+n^2}\\
\times&\sin\theta d\theta\wedge d\phi+\frac{4n^2[g(n^2-r^2)-er]}{(r^2+n^2)^2}\cos\theta\,dr\wedge d\phi.
\end{aligned}\ea
Performing the integration over a sphere $S_r$ of radius $r$ according to Eq. \eq{QeQm} , we can obtain the electric and magnetic charges inside this sphere
\ba\begin{aligned}
q_e(r)&=\frac{e(r^2-n^2)-4g r n^2}{r^2+n^2}\,,\\ q_m(r)&=\frac{2n[er+g(r^2-n^2)]}{r^2+n^2}\,. \label{qeqm}
\end{aligned}\ea
One can note that disparate with common results of the dyonic solutions, the charges vary with the radius of the sphere $S_r$, which can be easily understood since the sphere $S_r$ is not a close surface due to the existence of the Misner singularities. And two kinds of special values of these charges are respectively the asymptotic $(r\to\inf)$ charges:
\ba
Q_e=e\,,\ \ \ Q_m=2gn\,,
\ea
and the horizon $(r=r_\pm)$ charges:
\ba
Q_e^{(\pm)}=q_e(r_\pm)\,,\ \ \ Q_m^{(\pm)}=q_m(r_\pm)\,,
\ea
where $r_\pm$ denote the radius of the outer and inner horizons. We can note that the electromagnetic duality ($q_e^2\leftrightarrow q_m^2$) can be realized by $e\leftrightarrow -2ng, 2ng\leftrightarrow e$. By performing the conformal method \cite{AS}, the mass of this black hole is obtained by
\ba
M=m\,,
\ea
while the total angular momentum of the spacetime vanishes. According to the line element \eq{A}, the Killing vector of the horizon is manifestly given by
\begin{align}
k^a=\left(\frac{\pd}{\pd t}\right)^a\,.
\end{align}
And the temperature of these RN-type horizons are given by the surface gravity
\ba\begin{aligned}
T^{(\pm)}&=\frac{\k_\pm}{2\pi}=\frac{f'(r_\pm)}{4\pi}\\
&=\frac{1}{4\pi r_\pm}\left(1+\frac{3(r_\pm^2+n^2)}{l^2}-\frac{e^2+4n^2 g^2}{r_\pm^2+n^2}\right)\,.
\end{aligned}\ea
The entropies of these horizons can be identified  with the horizon area law:
\begin{align}
S^{(\pm)}=\frac{A_\pm}{4}=\pi\left(r_\pm^2+n^2\right)\,.
\end{align}

In order to obtain an expected feature of the thermodynamics in this spacetime, Ref. \cite{DK2} also introduced some quantities related to the Misner string, namely, the Misner potential $\y$ and Misner charge $N^{(\pm)}$:
\ba\begin{aligned}\label{Misnercharge}
\y&=\y^{(\pm)}=\frac{1}{8\p n}\,.\\
N^{(\pm)}&=-\frac{4n^3\pi}{r_\pm}\left[1+\frac{3(n^2-r^2_\pm)}{l^2}+\frac{(r_\pm^2-n^2)(e^2+4egr_\pm)}{(r^2_\pm+n^2)^2}\right. \\
&\qquad\qquad \left.-\frac{4n^2g^2(3r_\pm^2+n^2)}{(r_\pm^2+n^2)^2}\right]\,.
\end{aligned}\ea
Then, the first law of this black hole can be written as
\ba\begin{aligned}
\d M&=T^{(\pm)}\d S^{(\pm)}+V^{(\pm)}\d P\\
&+\f_e^{(\pm)} \d Q_e+\f_m^{(\pm)} \d Q_m^{(\pm)}+\y\d N^{(\pm)}\,,\\
\end{aligned}\ea
where
\ba\begin{aligned}
P&=-\frac{\L}{8\p}=\frac{3}{8\p l^2}\,,\\
V^{(\pm)}&=\frac{4}{3}\p r_\pm^3\lf(1+\frac{3n^2}{r_\pm^2}\right)
\end{aligned}\ea
are the thermodynamic pressure and thermodynamic volume of the outer and inner horizons,
\begin{align}\label{EMpotential}
\phi_e^{(\pm)}=\frac{er_\pm-2gn^2}{r^2_\pm+n^2}\,,\ \ \ \phi_m^{(\pm)}=\frac{n(2gr_\pm+e)}{r^2_\pm+n^2}\,
\end{align}
are the electric and magnetic potentials of the RN-type horizons.
\begin{figure*}
\centering
\includegraphics[width=0.95\textwidth]{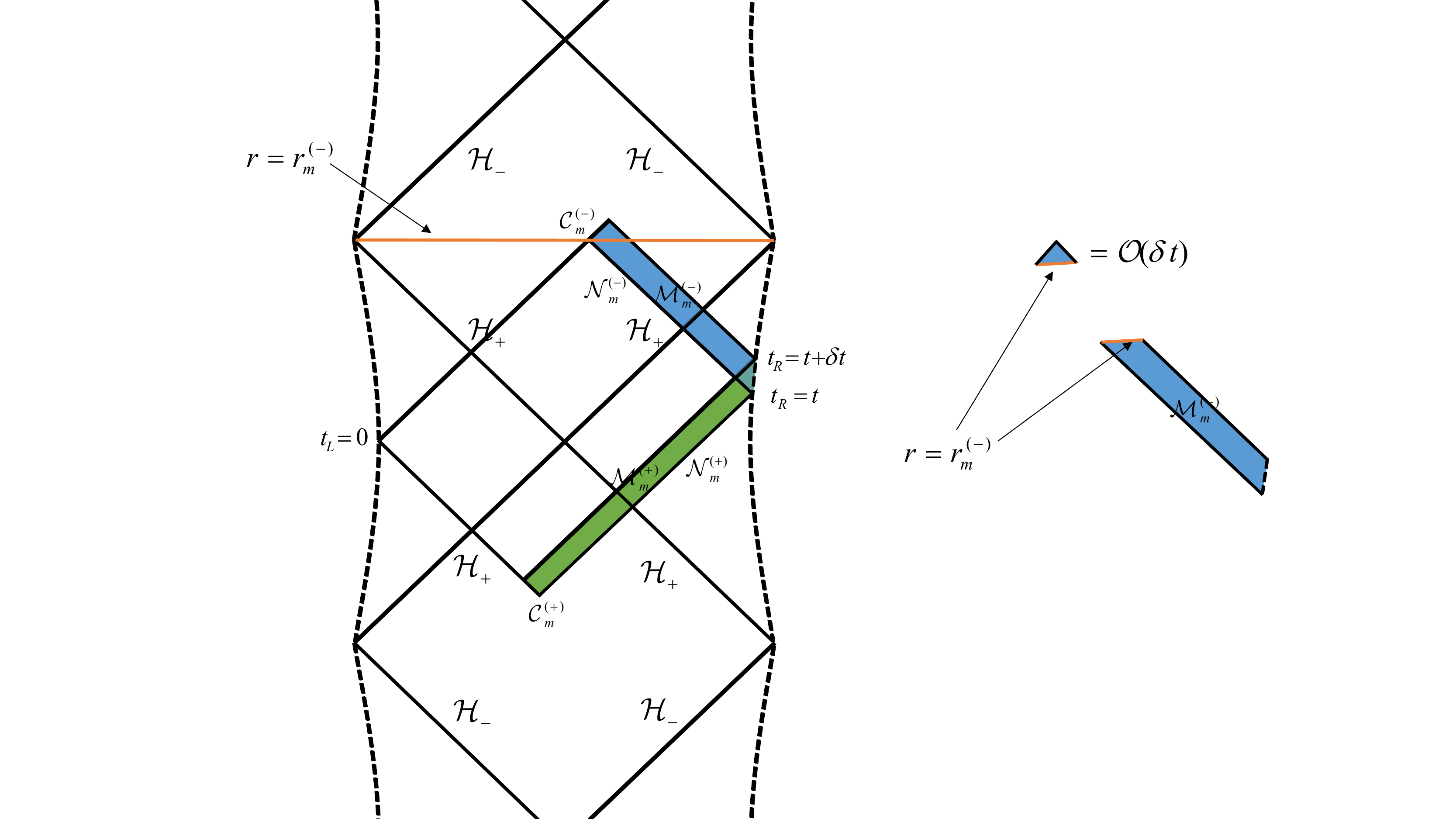}
\caption{Wheeler-DeWitt patch of a charged Taub-NUT-AdS black holes with two RN-type horizons, where the dashed lines denote the cut-off surface at asymptotic infinity, satisfying the asymptotic symmetries. In the left panel, we show that under the first-order approximation of $\d t$, the bulk region $\d \math{M}^{(-)}_m$ can be generated by the Killing vector $k^a$ through $\math{N}_m^{(-)}$.}\label{WDW}
\end{figure*}

\section{complexity growth rate in original CA conjecture}\label{3}
In this section, we use the ``complexity equals action'' conjecture to evaluate the complexity growth rate of the charged Taub-NUT-AdS black hole of general relativity coupled with a Maxwell field. As suggested by \cite{A3}, the CA complexity is equal to the full on-shell action on the WDW patch, which includes not only the bulk action but the surface terms, corner terms, and counterterms as well. According to \cite{A3}, the total action is given by
\ba\begin{aligned}\label{fullaction}
&I=I_\text{bulk}+\frac{1}{8\p}\int_{\math{B}}d^3x\sqrt{|h|}K
\pm\frac{1}{8\p}\int_{\math{C}}d^2x\sqrt{\s}\h\\
&+\frac{1}{8\p}\int_\math{N}d\l d^2\q \sqrt{\g} \k +\frac{1}{8\p}\int_{\math{N}}d\l d^2\q\sqrt{\g}\Q\ln\lf(\ell_{\text{ct}}\Q\rt)\,,
\end{aligned}\ea
where $K=\grad_a n^a$ is the trace of the extrinsic curvature, $\l$ is the parameter of the null generator $k^a$ on the null segment, $\k$ measures the failure of $\l$ to be an affine parameter which is derived from $k^a\grad_ak^b=\k k^b$, $\Q=\nabla_ak^a$ is the expansion scalar, and $\ell_\text{ct}$ is an arbitrary length scale. Here the surface term and joint term are introduced to make the variational principle well-posed. And the counterterm is added to ensure reparametrization invariance on the null boundaries.

Next, to evaluate the complexity growth rate in the CA context, we consider the change of the action within the WDW patch.  By considering the shift symmetry of this spacetime geometry, we can fix the left boundary time $t_L=0$ and only vary the right boundary time $t_R=t$. Then, the changes of the WDW patch of the charged Taub-NUT-AdS black holes with two RN-type horizons can be illustrated in the Penrose diagram \fig{WDW}. To regulate the divergence near the AdS boundary, a cut-off surface $r=r_\L$ is introduced. Since we only consider the time growth rate of the complexity, we can neglect the higher-order term of $\d t$. As illustrated in the right panel of \fig{WDW}, under the first-order approximation of $\d t$, the bulk region $\d \math{M}^{(\pm)}_m$ can be generated by the Killing vector $k^a$ through an asymptotic null hypersurface $\math{N}^{(\pm)}_m$ which terminates on the $2$-sphere $\math{C}_m^{(\pm)}$ with the radius $r=r_m^{(\pm)}$. Without loss of generality, we shall adopt the affine parameter for the null generator of the null surface; as a consequence, the surface term vanishes on all null boundaries. With these in mind, the change of the total action can be written as
\ba\begin{aligned}
\d I_\text{WDW}=I_{\d \math{M}^{(-)}_m}-I_{\d\math{M}^{(+)}_m}+\d I_{\math{C}_m^{(-)}}-\d I_{\math{C}_m^{(+)}}+\d I_\text{ct}\,.
\end{aligned}\nn\\\ea

\noindent\textbf{Bulk contributions}

We start by evaluating the contributions from the bulk action $\d \math{M}_m^{(-)}$ in the complexity growth rate. For simplification, we will neglect the index for the quantities in this region. For the bulk contribution from the gravitational action, we have
\ba\begin{aligned}\label{bulkgrav}
I_\text{grav}=\frac{\L}{8\p}\int_{\d \math{M}}\sqrt{-g}d^4x=\frac{\L \d t}{8\p}\int_{\math{N}}\star\bm{k}\,.
\end{aligned}\ea

Together with the equations of motion \eq{eom}, utilizing the facts that $k^a$ is a Killing vector and the sum of cyclic permutations of the last three indices in ${R^a}_{bcd}$ vanishes, we have the following identity:
\ba\begin{aligned}\label{grad}
\grad_a \grad^a k^b&=-{R^b}_ak^a\\
&=-\Lambda k^b-T^{ab}k_a\,,
\end{aligned}\ea

From the equation of motion $d\bm{G}=0$ of the electromagnetic field, we can see that the on-shell value of $\bm{G}$ is a closed $2$-form, which implies that there exists a $1$-form ${\bm B}$ such that ${\bm G}=d{\bm B}$. We can note that the new-defined vector potential $\bm{B}$ also allows a gauge freedom. Using the solution \eq{A}, it is not difficult to verify that this vector potential can be expressed as
\begin{align}\label{eqb}
{\bm B}=[p(r)-p_0]dt+2np(r)\cos\theta\,d\phi
\end{align}
with
\begin{align}
p(r)=\frac{4n^2gr-e(r^2-n^2)}{2n(r^2+n^2)}\,,
\end{align}
and arbitrary constant $p_0$ which should be determined when we fix the gauge. Substituting  Eq. \eq{Tab} into \eq{grad}, the second term becomes
\ba\begin{aligned}\label{Tk}
&T^{ab}k_a=k_a{F^a}_cF^{bc}+k_a{G^a}_cG^{bc}\\
&=F^{bc}k_a\grad^a A_c-F^{bc}k_a\grad_c A^a+G^{bc}k_a\grad^a B_c-G^{bc}k_a\grad_c B^a\\
&=-F^{bc}\grad_c \Phi_e+G^{bc}\grad_c \Phi_m\\
&=\grad_a\left(F^{ab}\Phi_e-G^{ab}\Phi_m\right)\,,
\end{aligned}\ea
where we have used the fact that Lie derivatives of vector potentials ${\bm A}$ and ${\bm B}$ along the vector $k^a$ vanish, namely ${\mathcal L}_{\bm k}\bm{A}=0$ and ${\mathcal L}_{\bm k}\bm{B}=0$, and denote the contractions of Killing vector $k^a$ with two vector potentials as
\begin{align}
\Phi_e=k^aA_a\,, \quad \Phi_m=-k^aB_a\,, \label{kaba}
\end{align}
which can be regarded as the electric and magnetic potentials of Maxwell field. And the counterparts of the inner and outer horizon can be defined as
\ba\begin{aligned}
\f_e^{(\pm)}&=\F_e(\inf)-\F_e(r_\pm)\,,\\
\f_m^{(\pm)}&=\F_m(\inf)-\F_m(r_\pm).\label{phiephim}
\end{aligned}\ea
From Eqs. (\ref{A}) and (\ref{eqb}), it is not difficult to see that they are actually the potentials as shown in Eq. \eq{EMpotential}.
Then, Eq. (\ref{grad}) together with (\ref{Tk}) yields
\begin{align}
\grad_a\left(\grad^a k^b+F^{ab}\Phi_e-G^{ab}\Phi_m\right)=-\L k^b\,.
\end{align}
Using the language of differential forms, we have
\begin{align}
2\L\star{\bm k}=d\left(\star d{\bm k} +2\Phi_e{\bm G}+2\Phi_m{\bm F}\right)\,.
\end{align}
Substituting above expression into \eq{bulkgrav}, the gravitational part \eq{bulkgrav} can be expressed as
\ba\begin{aligned}
I_\text{grav}=\frac{\d t}{16\p}\int_\math{N}d\left(\star d{\bm k} +2\Phi_e{\bm G}+2\Phi_m{\bm F}\right)\,.
\end{aligned}\ea
Next, we consider the bulk contribution from the electromagnetic field,
\ba\begin{aligned}\label{EMterm}
I_\text{EM}&=-\frac{1}{8\p}\int_{\d \math{M}}\bm{F}\wedge \bm{G}\\
&=-\frac{\d t}{8\p}\int_{\math{N}}k\.\lf(\bm{F}\wedge \bm{G}\rt)\\
&=-\frac{\d t}{8\p}\int_{\math{N}}\lf[(k\.d\bm{A})\wedge \bm{G}+\bm{F}\wedge (k\.d\bm{B})\rt]\\
&=\frac{\d t}{8\p}\int_{\math{N}}d\lf(\F_e\bm{G}-\F_m\bm{F}\rt)\,,\\
\end{aligned}\ea
where we also used $\math{L}_k\bm{A}=\math{L}_k\bm{B}=0$, as well as $d\bm{F}=d\bm{G}=0$ for the on-shell field $\bm{A}$. Combing these results, the bulk contribution becomes
\ba\begin{aligned}\label{bulkaction}
I_{\d \math{M}}&=I_\text{grav}+I_\text{EM}\\
&=\frac{\d t}{16\p}\int_\math{N}d\left(\star d{\bm k} +4\Phi_e{\bm G}\right)\\
&=\frac{\d t}{16\p}\int_{\pd\math{N}}\left(\star d{\bm k} +4\Phi_e{\bm G}\right)
\end{aligned}\ea
\begin{figure}
\centering
\includegraphics[width=0.5\textwidth]{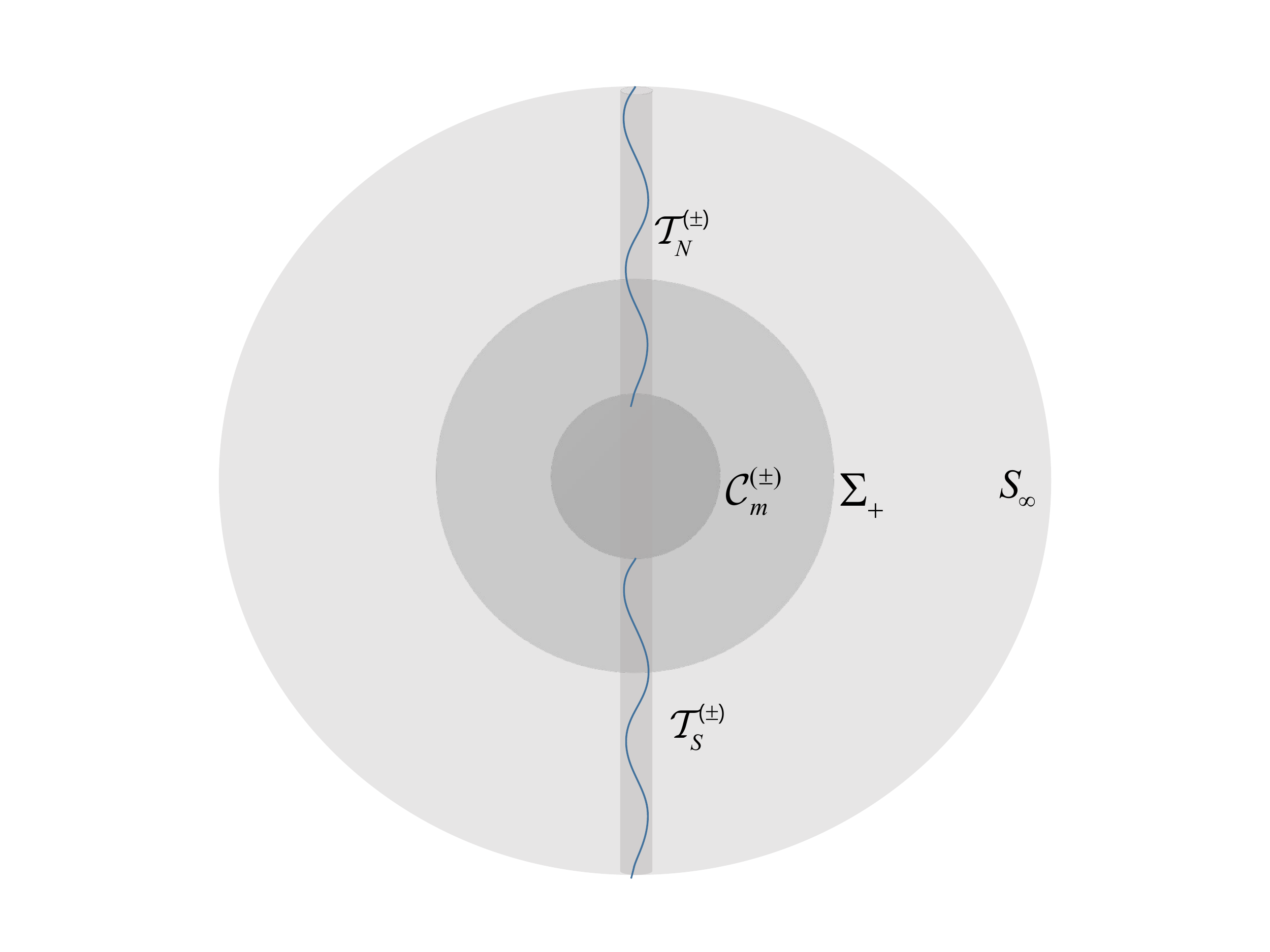}
\caption{{The null segment $\math{N}_m^{(\pm)}$ of WDW patch in charged Taub-NUT-AdS spacetimes: Misner tubes}, as a result of the effect of NUT parameter $n$, besides the standard boundaries $\math{C}_m^{(\pm)}$ and at infinity $S_\infty$, the two Misner tubes $\math{T}_N^{(\pm)}$ and $\math{T}_S^{(\pm)}$ of radius $\epsilon$ surrounding the symmetric strings are located at the north and south pole axes, $\cos\theta=\pm1$, which connected $\math{C}_m^{(\pm)}$ and $S_\inf$. Here $\S_\pm$ denote the portion on the horizon $\math{H}_\pm$.}\label{fig}
\end{figure}

For the normal black holes, the hypersurface $\math{N}$ is bounded by the spheres $\math{C}$ with the radius $r=r_m^{(-)}$ and $S_\inf$ at asymptotic infinity. However, when the Misner strings are present, there exists the Misner string singularities which locate at $\q=0$ and $\q=\pi$. Then, the decomposition will depend on the Misner strings (See \fig{fig}). Therefore, we also need to introduce two Misner string tubes $\math{T}_N$ and $\math{T}_S$ which are located at $\q=\epsilon$ and $\q=\pi-\epsilon$ with infinitesimal parameter $\epsilon$, respectively. After taking into account the orientation of the boundaries, we have
\begin{align}
\pd\math{N}=\math{T}_N+S_\inf-\math{T}_S-\math{C}\,.
\end{align}
Thus, the integral \eq{bulkaction} can be decomposed into the following integrations over several pieces of boundaries:
\ba\begin{aligned}\label{dM}
I_{\d \math{M}}&=\frac{\d t}{16\p}\int_{S_\inf}\star d\bm{k}-\frac{\d t}{16\p}\int_{\math{C}}\star d\bm{k}+\frac{\d t}{4\p}\int_{S_\inf} \F_e \bm{G}\\
&-\frac{\d t}{4\p}\int_{\math{C}}\F_e \bm{G}+\frac{\d t}{16\p}\int_{\math{T}_N}\left(\star d{\bm k} +4\Phi_e{\bm G}\right)\\
&-\frac{\d t}{16\p}\int_{\math{T}_S}\left(\star d{\bm k} +4\Phi_e{\bm G}\right)\,.
\end{aligned}\ea

By considering line element \eq{A} of this spacetime, we evaluate the Hodge dual of $k^a$, namely
\begin{align}
\star d{\bm k}=&-\frac{2nf(r)}{r^2+n^2}\,dr\wedge(dt+2n\cos\theta\,d\phi)\n\\
&-\sin\theta(r^2+n^2)f'(r)\,d\theta\wedge d\phi.
\end{align}
Then, we can further obtain
\ba\begin{aligned}
\int_{\math{C}}\star d\bm{k}&=-4 \p (r^2+n^2)f'(r)
\end{aligned}\ea
Considering the fact that $r\to r_-$ at the late times, we have
\ba
-\frac{1}{16\p}\lim_{t\to\inf}\int_{\math{C}}\star d\bm{k}=TS\,.
\ea
At the same time, we can also obtain
\ba\begin{aligned}
&\int_{\math{T}_N}\star d{\bm k}=-\int_{\math{T}_S}\star d{\bm k}\\
&=-\frac{8\pi n^2r_\L}{l^2}-\frac{2\pi^2(e^2+4g^2n^2)}{n}+\frac{8\pi n^2r}{l^2}\\
&+\frac{4\pi\left[2mn^2+(e^2+4g^2n^2-2n^2)r\right]}{r^2+n^2}-\frac{32\pi n^4r}{l^2(r^2+n^2)}\\
&+\frac{4\pi(e^2+4g^2n^2)\arctan\left(r/n\right)}{n}\,.
\end{aligned}\ea

Finally, we evaluate the contributions from the electromagnetic field. According to the above calculation, we can see that our final result \eq{dM} is independent on the gauge choice of electromagnetic potentials. For later convenience, here we choose the gauge such that $\F_e(r_-)=0$, i.e.,
\ba\begin{aligned}
\F_e(r)=\F_e^{(-)}(r)=\frac{er_--2n^2 g}{n^2+r^2_-}-\frac{er-2n^2 g}{n^2+r^2}\,.
\end{aligned}\ea
Then, we can further obtain
\ba\begin{aligned}
&\int_{\math{T}_N}4\Phi_e{\bm G}=-\int_{\math{T}_S}4\Phi_e{\bm G}\\
&=\frac{2\pi^2(e^2+4g^2n^2)}{n}-\frac{4\pi(e^2+4g^2n^2)\arctan\left(r_h/n\right)}{n}\\
&+\frac{8n^2\p\lf[e^2r-4n^2g(e+g r)\rt]}{(n^2+r^2)^2}-\frac{16\p n^2\f_e^{(-)}(e+2 g r)}{n^2+r^2}\\
&-\frac{4\p[e^2 r-4n^2g(e-g r)]}{n^2+r^2}\,.\\
\end{aligned}\ea
From these results, Eq. \eq{dM} can be expressed by
\ba\begin{aligned}
I_{\d \math{M}}/\d t&=\frac{(r^2+n^2)f'(r)}{4}+\f_e^{(-)} Q_e-\F_e^{(-)}(r)q_e(r)\\
&+K^{(-)}(r)+\frac{1}{16\p}\int_{S_\inf}\star d\bm{k}
\end{aligned}\ea
where we have denoted
\ba\begin{aligned}
&K^{(\pm)}(r)=\frac{1}{8\p}\int_{\math{T}_N^{(\pm)}}\left(\star d{\bm k} +4\Phi_e^{(\pm)}{\bm G}\right)\\
&=\frac{n^2r(r^2-3n^2)}{l^2(n^2+r^2)}+\frac{n^2 r(e^2+2e g r-r^2)}{(n^2+r^2)^2}+\frac{m n^2}{n^2+r^2}\\
&-\frac{n^4[r+2g(e+2g r)]}{(n^2+r^2)^2}-\frac{2 n^2\f_e^{(\pm)}(e+2 g r)}{n^2+r^2}-\frac{n^2 r_\L}{l^2}\,,
\end{aligned}\ea
with
\ba\begin{aligned}
\F_e^{(+)}(r)=\frac{er_+-2n^2 g}{n^2+r^2_+}-\frac{er-2n^2 g}{n^2+r^2}
\end{aligned}\ea
such that $\F_e^{(+)}(r_+)=0$. With similar calculation, we can also obtain the contributions from $\d M_m^{(+)}$ and the growth rate of the bulk action can be read off
\ba\begin{aligned}
\frac{d I_\text{bulk}}{d t}=\lf[\f_e^{(h)}Q_e+K^{(h)}-\F_e^{(h)}q_e+\frac{(r^2+n^2)f'(r)}{4}\rt]^{\c_-}_{\c_+}\,,
\end{aligned}\nn\\\ea
where we denote $\c_\pm\equiv\{h=\pm,\, r=r_m^{(\pm)}\}$. By comparing the expressions of the Misner potential and charge \eq{Misnercharge}, together with the fact $r_m^{(\pm)}\to r_\pm$ at late times, we can easily verify
\ba
K^{(-)}(r_-)-K^{(+)}(r_+)=\y \lf(N^{(-)}-N^{(+)}\rt)\,.
\ea
Together with the fact $\F_e^{(\pm)}(r_\pm)=0$, the late-time growth rate of the bulk action can be expressed as
\ba\begin{aligned}
\lim_{t\to\inf}&\frac{d I_\text{bulk}}{d t}=T^{(-)}S^{(-)}-T^{(+)}S^{(+)}\\
&+\lf(\f_e^{(-)}-\f_e^{(+)}\rt) Q_e+\y\lf( N^{(-)}-N^{(+)}\rt)\,.
\end{aligned}\ea

\noindent \textbf{Joint contributions}

We next calculate the joint contributions from meeting points $\math{C}_m^{(\pm)}$. From the line element \eq{A}, it is not hard to find that
\ba\begin{aligned}\label{k1k2}
k_1^a&=\frac{1}{f(r)}\lf(\frac{\pd}{\pd t}\rt)^a+\lf(\frac{\pd}{\pd r}\rt)^a\,,\\
k_2^a&=-\frac{1}{f(r)}\lf(\frac{\pd}{\pd t}\rt)^a+\lf(\frac{\pd}{\pd r}\rt)^a\,
\end{aligned}\ea
are the affinely null generator of the past right and past left null boundaries separately, which means that the null surface term will vanishes under this choice. Then, we can obtain $k_1\cdot k_2=2/f$. By using the transformation parameter $\h=\ln(|k_1\cdot k_2|/2)$, we have
\ba\begin{aligned}
I_{\math{C}_m^{(+)}}=-\frac{1}{2}\lf[(r_m^{(+)})^2+n^2\rt]\ln \lf[-f(r_m^{(+)})\rt]\,.
\end{aligned}\ea
According to \eq{k1k2}, the integral curves of the two generators can be expressed as
$
t(\l)=t+r^\star (\l)\,,\ r(\l)=\l\,,\ \q(\l)=\q\,, \ \f(\l)=\f
$
and
$t(\l)=-r^\star (\l)\,,\ r(\l)=\l\,,\ \q(\l)=\q\,, \ \f(\l)=\f$, separately. Here we defined the tortoise coordinates,
\ba\begin{aligned}\label{rstar}
r^\star(r)&=-\int_r^\inf \frac{dr}{f(r)}\,,
\end{aligned}\ea
where this range of integration is chosen to make that the coordinate satisfy the boundary condition $\lim_{r\to \inf}r_\star (r)=0$.
According to \cite{A4}, it can be written as
\ba\begin{aligned}
r^\star(r)&=\frac{\ln(|r-r_+|/r)}{g(r_+)(r_+-r_-)}-\frac{\ln(|r-r_-|/r)}{g(r_-)(r_+-r_-)}\\
&-\frac{1}{r_+-r_-}\int_{r}^{\inf} G(r)dr
\end{aligned}\ea
with
\ba\begin{aligned}
g(r)&=\frac{f(r)}{(r-r_+)(r-r_-)}\,,\\
G(r)&=\frac{g(r_+)r-g(r)r_+}{g(r_+)g(r)r(r-r_+)}
-\frac{g(r_-)r-g(r)r_-}{g(r_-)g(r)r(r-r_-)}\,.
\end{aligned}\ea
By using the integral curves of the generator, the radius of the meeting point $\math{C}_m^{(+)}$ can be obtained by
\ba\label{rmp}
r^\star(r_m^{(+)})=-\frac{t}{2}\,.
\ea
Then, the change rate of this dynamical point can be read off
\ba\label{drp}
\frac{d r_m^{(+)}}{d t}=-\frac{1}{2}f(r_m^{(+)})\,.
\ea
The time derivative of the joint action can be expressed by
\ba\begin{aligned}
\frac{d I_{\math{C}_m^{(+)}}}{dt}=\left.\lf[\frac{(r^2+n^2)f'(r)}{4}+\frac{r f(r)}{2}\ln \lf(-f(r)\rt)\rt]\right|_{r=r_m^{(+)}}\,.
\end{aligned}\ea
With similar consideration, we can further obtain the contribution from the joint $\math{C}_m^{(-)}$. And the growth rate of the joint actions can be written as
\ba\begin{aligned}
\frac{d I_\text{joint}}{dt}= \lf[\frac{(r^2+n^2)f'(r)}{4}+\frac{r f(r)}{2}\ln \lf(-f(r)\rt)\rt]^{r_m^{(+)}}_{r_m^{(-)}}\,,
\end{aligned}\ea
where we used the ralation
\ba\label{rmm}
r^\star(r_m^{(-)})=\frac{t}{2}\,.
\ea
At the late times, we have
\ba\begin{aligned}
\lim_{t\to \inf}\frac{d I_\text{joint}}{dt}=T^{(+)}S^{(+)}-T^{(-)}S^{(-)}\,.
\end{aligned}\ea

\noindent\textbf{Counterterm contributions}

Finally, we consider the counterterm contributions. We first evaluate the contribution which comes from the past right null boundary with $\l$ as the affine parameter, i.e., $k_1^{a}=\lf(\pd/\pd \l\rt)^a$ in \eq{k1k2}, which gives rise to the expansion $\Q=2r/(n^2+r^2)$. Then, the counterterm of the past right null segment can be shown as
\ba\begin{aligned}
I_\text{ct}^\text{(pr)}=\int_{r_m^{(+)}}^{r_\L}d\l r(\l)\ln\lf(\frac{2r(\l)\ell_\text{ct}}{r(\l)^2+n^2}\rt)\,,
\end{aligned}\ea
By considering that $r(\l)=\l$, the rate of this counterterm becomes
\ba
\frac{d I_\text{ct}^\text{(pr)}}{dt}=\frac{1}{2} r f(r)\ln\lf(\frac{2r\ell_\text{ct}}{r^2+n^2}\rt)\,.
\ea
Again, we can obtain the counterterm contributions from other segments and the final result is given by
\ba
\frac{d I_\text{ct}}{dt}=\lf[r f(r)\ln\lf(\frac{2r\ell_\text{ct}}{r^2+n^2}\rt)\rt]^{r_m^{(+)}}_{r_m^{(-)}}\,.
\ea

\noindent\textbf{Complexity growth}

By summing all the previous results, one can further obtain
\ba\begin{aligned}
&\frac{d C_A}{d t}=\frac{1}{\p \hbar}\\
&\times\lf[K^{(h)}+\f_e^{(h)}Q_e-\F_e^{(h)}q_e-\frac{1}{2}rf\ln\lf(\frac{4r^2\ell_\text{ct}^2|f|}{(r^2+n^2)^2}\rt)\rt]^{\c_-}_{\c_+}
\end{aligned}\nn\\\ea
\begin{figure}
\centering
\includegraphics[width=0.45\textwidth]{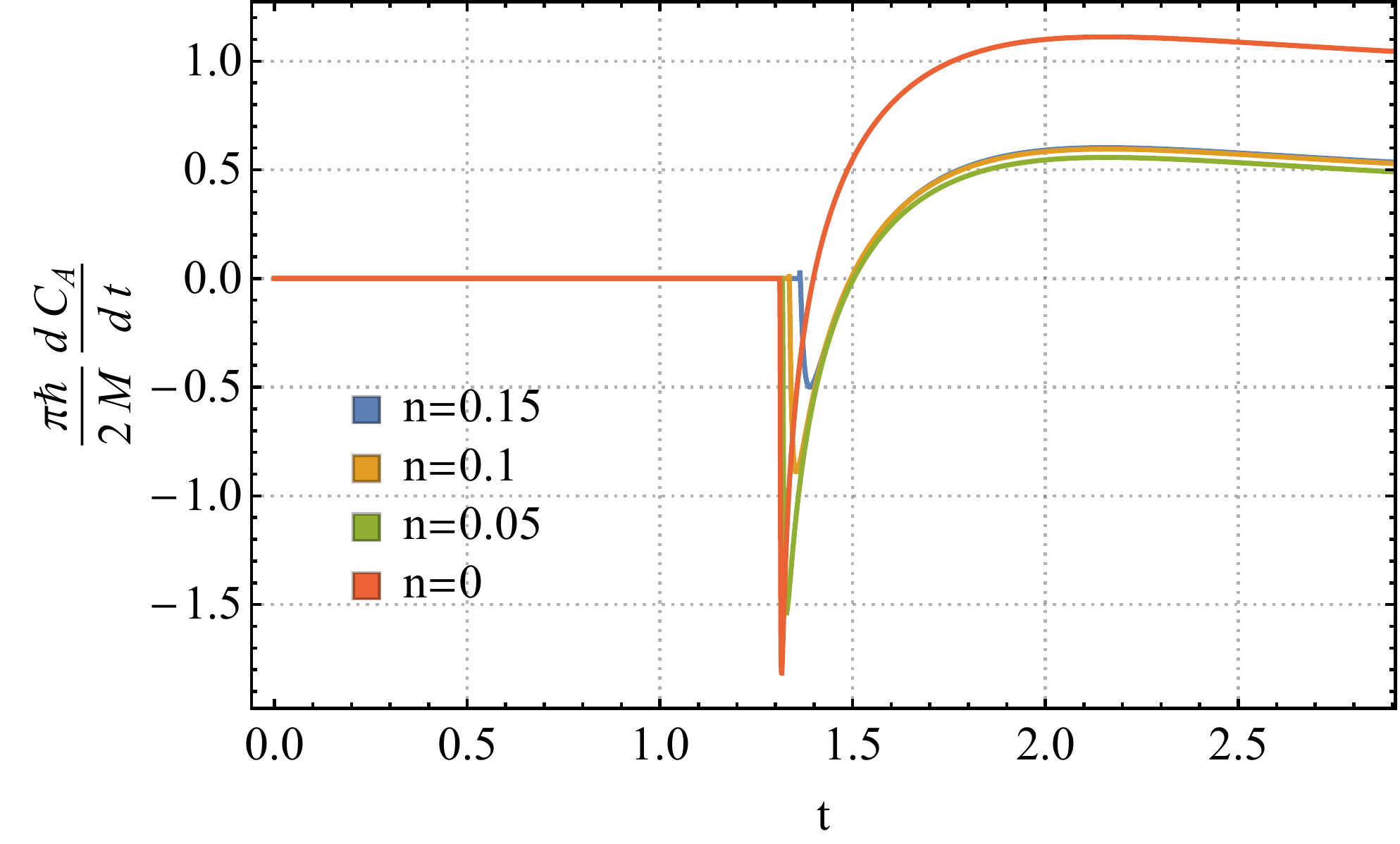}
\caption{The time dependence of the complexity growth rate in the context of original CA conjecture with $m=1, l=1, \ell_\text{ct}=0.1,e=0.2$, and $g=0.2$.}\label{dcdtf1}
\end{figure}

In \fig{dcdtf1}, we show the time-dependence of the complexity growth rate with the original CA conjecture. This figure shows a similar behavior with the case of the RN-AdS black hole in \cite{A4} in which the late-time value is approached above. Moreover, we can see that
for the case with small NUT and electromagnetic charges, the Lloyd's bound will be violated when we consider the full-time evolution of the complexity. This can be easily understood since the charged Taub-NUT-AdS black hole will return to the RN-AdS black hole and the CA complexity in RN-AdS can also violate the Lloyd's bound in some small charge cases \cite{A4}.

Finally, we consider the late time result of this holographic complexity. Summing above results, it can be written as
\ba\label{dCAdt1}\begin{aligned}
\lim_{t\to\inf}\frac{d C_A}{dt}=\frac{1}{\p \hbar}\lf[\lf(\f_e^{(-)}-\f_e^{(+)}\rt) Q_e+\y \lf(N^{(-)}-N^{(+)}\rt)\rt]\,.
\end{aligned}\nn\\\ea

We can note that in this case, the Misner string plays an important role in the holographic complexity. When the Misner string exists, the late-time complexity growth rate is dependent on the Misner potential and Misner charge which is introduced in Ref. \cite{DK2} to obtain the expected features of the thermodynamics in the charged Taub-NUT-AdS black holes. Moreover, differing from the normal charged black holes, where the charges appeared in the late-time rate of the CA complexity only depends on their values on the horizons, the electric charge here is the total charge of the black hole. These results imply that the holographic complexity can exactly reflect the nontrivial topological of this spacetime, such as the existence of the Misner string. Moreover, similar to the dyonic black hole, the late-time growth rate of this original CA conjecture only depends on the electric charge. And one can verify that this result also violates the electromagnetic duality of the Maxwell theory. It was recently proposed in \cite{Goto:2018iay} that this electromagnetic duality can be restored by adding the Maxwell boundary terms. Therefore, in the next section, we would like to study the CA complexity with Maxwell boundary terms.

\begin{figure*}
\centering
\includegraphics[width=0.47\textwidth]{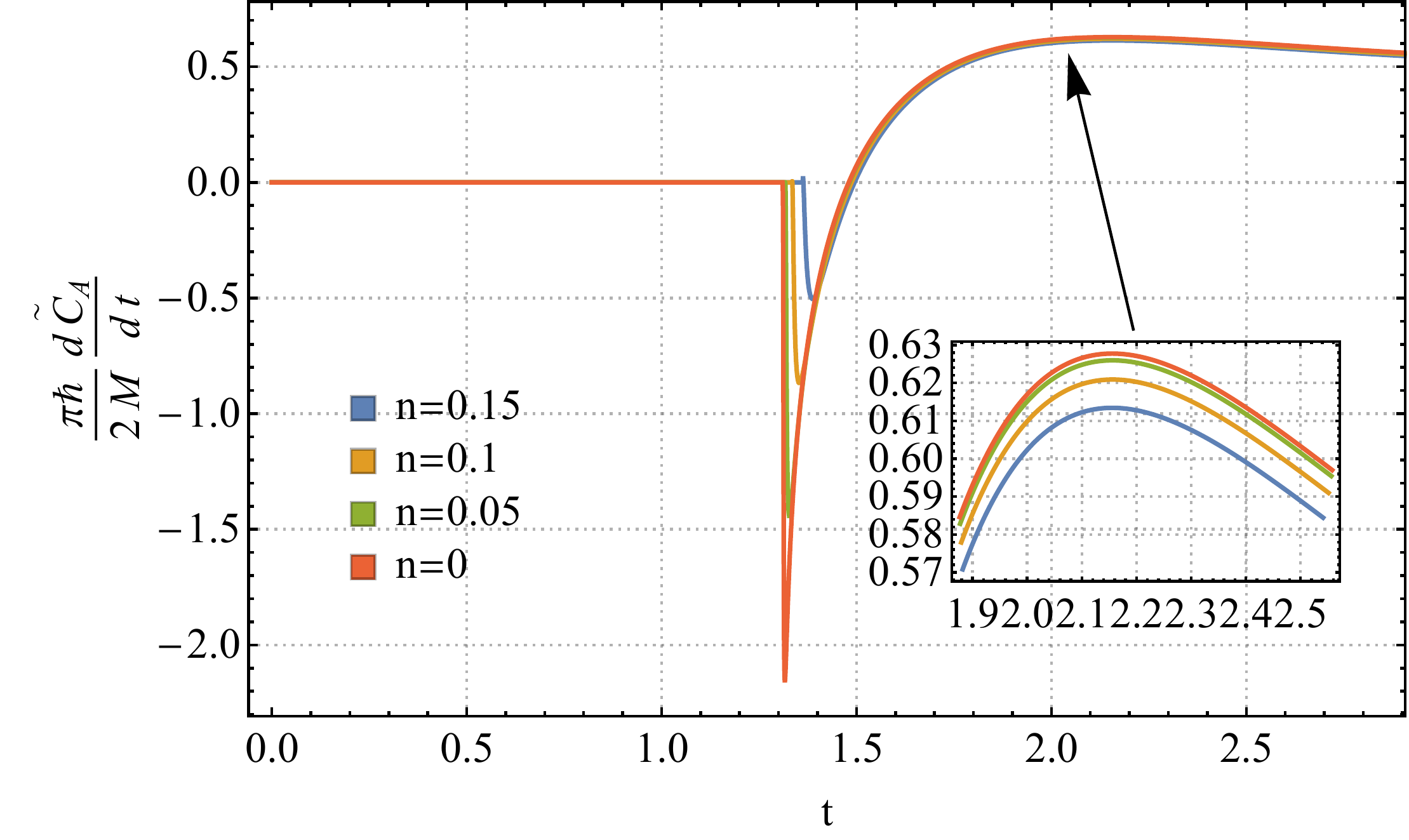}
\includegraphics[width=0.47\textwidth]{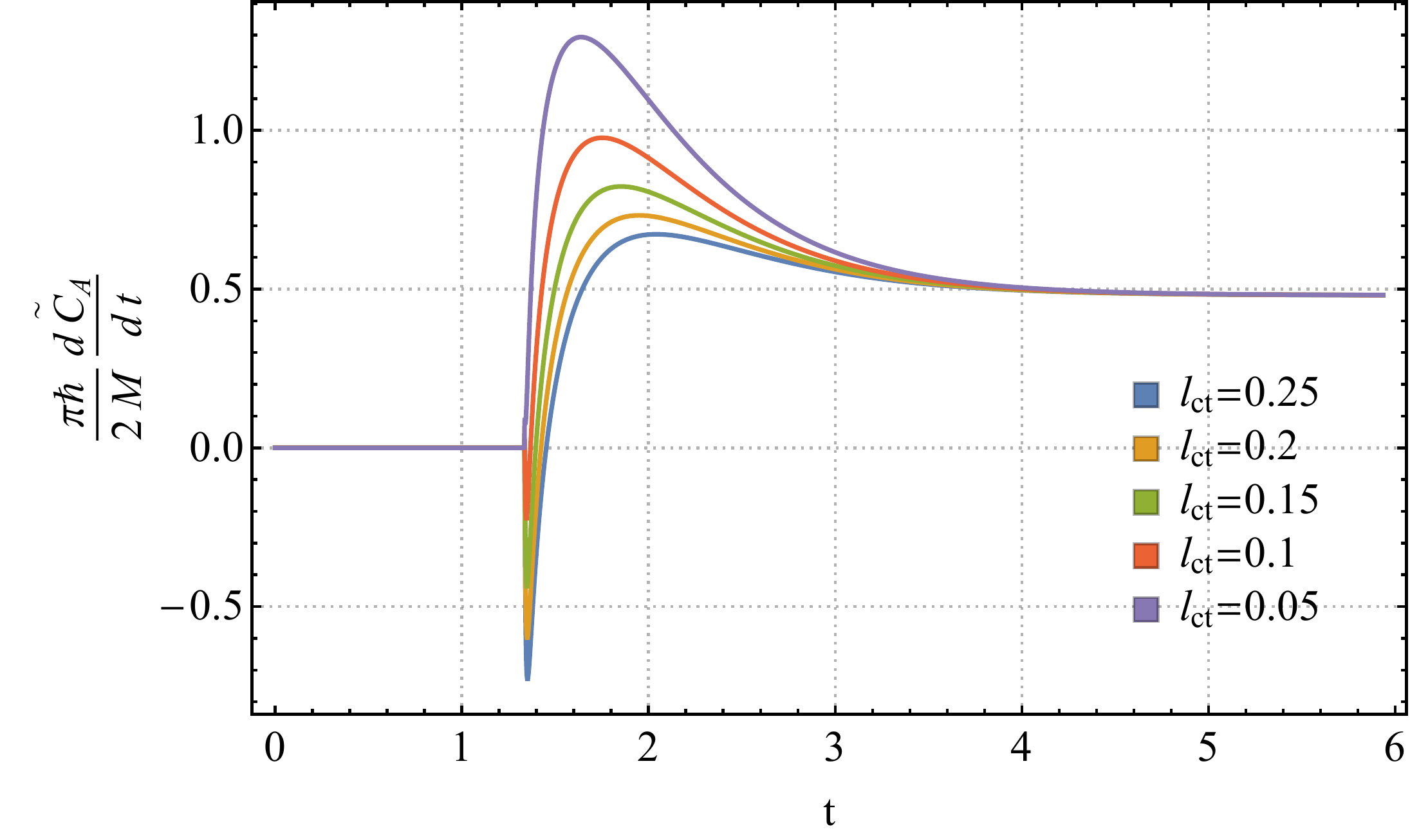}
\caption{The time dependence of the complexity growth rate in the context of CA conjecture with Maxwell boundary term. In the left panel, we set $m=1, l=1, \ell_\text{ct}=0.1, e=0.2, g=0.2$ and change the NUT parameter $n$. In the right panel, we set $m=1, l=1,e=0.2, g=0.2, n=0.1$ and change scale length $\ell_\text{ct}$. }\label{dcdtf2}
\end{figure*}
\section{Complexity growth rate with Maxwell boundary term}\label{4}

In \cite{Goto:2018iay}, the author discussed the CA complexity of the dyonic black hole in Einstein-Maxwell gravity. They found that the  original CA complexity in this case also violates the electromagnetic duality, and for the purely magnetic case, the late-time complexity growth rate always vanishes. In order to obtain the expected features of the CA duality, they suggested that the complexity should be modified by adding the additional Maxwell boundary term,
\ba
I_{\m \rm Q}=\frac{\g}{4\p}\int_{\pd \math{M}}\bm{G}\wedge \bm{A}\,,
\ea
where $\g$ is a free parameter.  Unlike the Gibbons-Hawking surface term, this boundary term was not required by the variation principle, and it only gives different boundary conditions of the electromagnetic field. Then, the full action is given by
\ba
I_\text{full}=I+I_{\m\text{Q}}\,.
\ea

If the electromagnetic field satisfies the equation of motion $d\bm{G}=0$, using the Stokes' theory, this boundary term is equivalent to
\ba\label{ImQM}
I_{\m\text{Q}}=\int_\math{M} \bm{L}_{\m\text{Q}}
\ea
with
\ba\label{LmQ}
\bm{L}_{\m\text{Q}}=\frac{\g}{4\p}\bm{G}\wedge \bm{F}\,.
\ea
With the similar calculation as the EM bulk term \eq{EMterm}, the contributions from the Maxwell boundary term to the late-time complexity growth rate can be shown as
\ba\begin{aligned}
&\d I_{\m\text{Q}}=-\frac{\g\d t}{4\p}\lf[\int_{\math{N}_m^{(-)}}d\lf(\F_e^{(-)}\bm{G}-\F_m \bm{F}\rt)\right.\\
&\left.-\int_{\math{N}_m^{(+)}}d\lf(\F_e^{(+)}\bm{G}-\F_m \bm{F}\rt)\rt]\,.
\end{aligned}\ea

Without loss of generality, here we choose the gauge such that $\F_m (\inf)=0$, i.e., we have $\F_m(r_\pm)=-\f_m^{(\pm)}$ and
\ba
\F_m(r)=-\frac{n(e+2gr)}{n^2+r^2}\,.
\ea
Using the Eqs. (\ref{expF}) and (\ref{expG}), we can further obtain
\ba\begin{aligned}
&\D_q^{(\pm)}(r_m^{(\pm)})=\frac{1}{2\p}\int_{\math{T}_N^{(\pm)}}\lf(\F_e^{(\pm)}\bm{G}-\F_m\bm{F}\rt)\\
&=\left.\frac{2n^2(r_h-r)(e+2g r)\lf[err_h-n^2(e+2gr_h+2gr)\rt]}{(n^2+r_h^2)(n^2+r^2)^2}\right|_{\c_\pm}\,.
\end{aligned}\nn\\\ea
Then, we have
\ba\begin{aligned}
\frac{d I_{\m\text{Q}}}{d t}=\g\lf[\f_e^{(\pm)} Q_e-\F_e^{(\pm)}q_e+\F_m q_m+\D_q^{(h)}\rt]^{\c_+}_{\c_-}\,.\\
\end{aligned}\ea

The complexity growth rate with this boundary term can be expressed by
\ba\begin{aligned}
&\frac{d \tilde{C}_A}{d t}=\frac{1}{\p \hbar}\lf[K^{(h)}+(1-\g)\lf(\f_e^{(h)}Q_e-\F_e^{(h)}q_e\rt)\right.\\
&\left.+\g\F_m q_m+\g\D_q^{(h)}-\frac{1}{2}rf\ln\lf(\frac{4r^2\ell_\text{ct}^2|f|}{(r^2+n^2)^2}\rt)\rt]^{\c_-}_{\c_+}\,.
\end{aligned}\ea

At the late times, it becomes
\ba\label{dCAdt2}\begin{aligned}
&\lim_{t\to\inf}\frac{d \tilde{C}_A}{dt}=\frac{1}{\p \hbar}\lf[(1-\g)\lf(\f_e^{(-)}-\f_e^{(+)}\rt) Q_e\right.\\
&\left.\g \lf(\f_m^{(-)}Q_m^{(-)}-\f_m^{(+)}Q_m^{(+)}\rt)+\y \lf(N^{(-)}-N^{(+)}\rt)\rt]\,.
\end{aligned}\ea

We can note that in the charged Taub-NUT-AdS black hole, the Maxwell boundary term only affects the proportion between the electric and magnetic terms, and it does not contribute to the part of the Misner term at the late times. Moreover, we can see that different with the electric charge, here the magnetic charge is determined by the inner and outer horizons. It is not difficult to verify that the case with $\g=1/2$ also gives the result which satisfies the electromagnetic duality. Then, the complexity growth rate becomes
\ba\begin{aligned}
\frac{d \tilde{C}_A}{dt}&=\lf[\frac{2n^2r^3-6n^4r+l^2(2m n^2+e^2 r-2n^2r+4n^2g^2r)}{2\p \hbar l^2(n^2+r^2)}\right.\\
&-\left.\frac{rf(r)}{2\p\hbar}\ln\lf(-\frac{4r^2\ell_\text{ct}^2f(r)}{(r^2+n^2)^2}\rt)\rt]^{r_m^{(-)}}_{r_m^{(+)}}\,.
\end{aligned}\ea
Then, its late-time limit can be shown as
\ba\begin{aligned}
\frac{\p \hbar}{2M}\lim_{t\to\inf}\frac{d \tilde{C}_A}{d t}&=\frac{(r_+-r_-)(l^2+3n^2+r_-^2+r_-r_++r_+^2)}{2(r_++r_-)(l^2+6n^2+r_-^2+r_+^2)}\\
&<\frac{1}{2}\,,
\end{aligned}\ea
which is much less than the Llyod's bound. In \fig{dcdtf2}, we show the time-dependence of the complexity growth rate in this modified CA conjecture with $\g=1/2$. This figure also shows a similar behavior with the case of the RN-AdS black hole. However, from the right panel of \fig{dcdtf2}, we can see that the maximal value of the complexity growth rate will decrease as the length scale $\ell_\text{ct}$ decreases and finally approach the late-time value. Therefore, in this case, the Lloyd's bound can always be satisfied as long as we choose a proper length scale $\ell_\text{ct}$, since the late-time value is much lower than the Lloyd's bound in this case.

\section{Conclusion}\label{5}

In this paper, we investigated the holographic complexity in the charged Taub-NUT-AdS black holes with two RN-type horizons present in the Einstein-Maxwell gravity. Differing from the normal black holes, the Misner string will affect the topology of the spacetime geometry. In the context of holography, this nontrivial topology also plays an important role in the holographic complexity. For the normal black holes, the late-time rate of the complexity only depends on the quantities on the outer and inner RN-type horizons. However, here it is also determined by the quantities on the Misner string, i.e., the Misner potential $\y$ and Misner charge $N^{(\pm)}$. Besides, we also found that the electric charge appeared in the late-time rate is the total charge of this black hole, while as indicated by the previous literature \cite{A4,Pan,A2,Guo:2017rul,WYY,Jiang1}, it is only the charges on the horizon for the normal black hole case.  These results imply that the holographic complexity can exactly reflect the non-trivial topological of this spacetime. Moreover, similar to the dyonic black hole, the late-time growth rate of this original CA conjecture only depends on the electric charge, and it violates the electromagnetic duality of the Maxwell theory. It was recently proposed in Ref. \cite{Goto:2018iay} that this electromagnetic duality can be restored by adding the Maxwell boundary terms. Therefore, in Sec. \ref{4}, we also studied the CA complexity with the additional Maxwell boundary terms. By adding the Maxwell boundary term, this late-time rate would become sensitive to the magnetic charge. And the special choice $\g=1/2$ can also make our final result satisfies the electromagnetic duality. Finally, we found that this additional term only changes the proportion between the electric and magnetic charges, and it does not affect the Misner term. Finally, we studied the time-dependence of the complexity growth rate with or without Maxwell boundary term and found that they share similar behaviors with that in RN-AdS black holes. For the special case with $\g=1/2$, the Lloyd's bound can always be satisfied as long as we choose some proper scale length $\ell_\text{ct}$.

\section*{acknowledgements}
This research was supported by National Natural Science Foundation of China (NSFC) with Grants No. 11375026, No. 11675015,
the Cultivating Program of Excellent Innovation Team of Chengdu University of Technology (Grant No. KYTD201704), the Cultivating Program of Middle-aged Backbone Teachers of Chengdu University of Technology (Grant No. 10912-2019KYGG01511), the Open Research Fund of Computational Physics Key Laboratory of Sichuan Province, Yibin University (Grant No. JSWL2018KFZ01)

\end{document}